\documentclass[twocolumn,aps,prl]{revtex4}
\usepackage{graphicx}

\newcommand{\rmssigma}{\sigma_{\mathrm{rms}}}

\begin{document}
\pagestyle{empty} 
\title{A multiscale Molecular Dynamics approach to Contact Mechanics}

\author{C. Yang,
        U. Tartaglino\footnote{
          Present address: DEMOCRITOS National Simulation Center,
          Via Beirut 2-4, I-34014 Trieste, Italy;
          e-mail: tartagli@sissa.it}
        and B.N.J. Persson}
\affiliation{IFF, FZ-J\"ulich, 52425 J\"ulich, Germany}

\begin{abstract}

The friction and adhesion between elastic bodies are strongly
influenced by the roughness of the surfaces in contact. 
Here we develop a multiscale molecular dynamics approach to 
contact mechanics, 
which can be used also when the surfaces have 
roughness on many different length-scales, e.g.,
for self affine fractal surfaces.
As an illustration we consider the contact between randomly rough surfaces,
and show that the contact area varies linearly with the load for 
small load.
We also analyze the contact morphology and the pressure distribution 
at different magnification, both with and without adhesion. 
The calculations are compared with analytical contact mechanics 
models based on continuum mechanics.

\end{abstract}
\maketitle


\vskip 0.5cm

{\bf 1. Introduction}

Adhesion and friction between solid surfaces are common phenomenons
in nature and of extreme importance 
in biology and technology. Most surfaces of solids have roughness on
many different length scales\cite{Krim,Valbusa}, 
and it is usually necessary to consider many decades in length
scale when describing the contact between solids\cite{P3}. 
This makes it very hard to describe accurately the contact mechanics between 
macroscopic solids using
computer simulation methods, e.g., atomistic molecular dynamics, or finite element 
calculations based on continuum mechanics.

Consider a solid with a nominally flat surface. Let $x,y,z$ be a coordinate system with the
$x,y$ plane parallel to the surface plane.
Assume that $z=h({\bf x})$ describe the surface height profile, where ${\bf x} = (x,y)$
is the position vector within the surface plane. The most important property characterizing 
a randomly rough surface is the surface roughness 
power spectrum $C({\bf q})$ defined by\cite{P3,Persson_JCP2001}
\begin{equation}
 \label{powerspectrum}
 C({\bf q}) = {1\over (2\pi )^2}
              \int d^2x \ \langle h({\bf x})h({\bf 0})\rangle 
              e^{i{\bf q}\cdot {\bf x}}.
\end{equation}
Here $\langle ... \rangle$ stands for ensemble average and we 
have assumed that $h({\bf x})$ is measured from the average surface plane so that 
$\langle h \rangle = 0$.
In what follows we will assume that
the statistical properties of the surface are isotropic, in which case $C(q)$ will only
depend on the magnitude $q=|{\bf q}|$ of the wave vector ${\bf q}$. 

\begin{figure}
\includegraphics[width=0.35\textwidth]{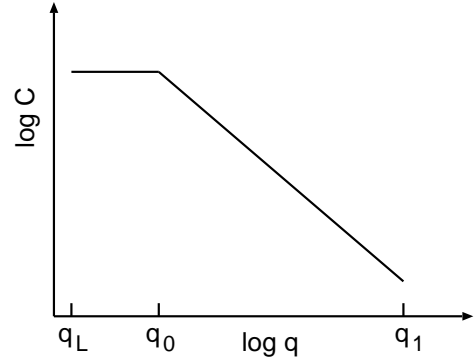} 
\caption{\label{Cq1}
Surface roughness power spectrum of a surface which is self affine fractal for
$q_1>q>q_0$. The long-distance roll-off wave vector $q_0$ and the short distance cut-off
wave vector $q_1$ depend on the system under consideration. The slope of
the ${\rm log} C-{\rm log}q$ relation for $q > q_0$ determines the fractal
exponent of the surface. The lateral size $L$ of the surface (or of the studied surface region)
determines the smallest possible wave vector $q_L=2\pi /L$.}
\end{figure}

Many surfaces tend to be nearly self-affine fractal. A self-affine fractal
surface has the property that if part of the surface is magnified, with a magnification
which in general is appropriately different in the perpendicular direction to the surface as compared
to the lateral directions, then the surface ``looks the same'', i.e., the statistical
properties of the surface are invariant under the scale transformation\cite{P3}.
For a self-affine
surface the power spectrum has the power-law behavior
\[
 C(q) \sim q^{-2(H+1)},
\]
where the Hurst exponent $H$ is related to the fractal dimension $D_{\rm f}$ of the surface via
$H=3-D_{\rm f}$. Of course, for real surfaces this relation only holds in some finite
wave vector region $q_0 < q < q_1$, and in a typical case $C(q)$ has the form shown
in Fig.~\ref{Cq1}. Note that in many cases there is a roll-off wavevector $q_0$ below which
$C(q)$ is approximately constant. 

\begin{figure}[htb]
   \includegraphics[width=0.3\textwidth]{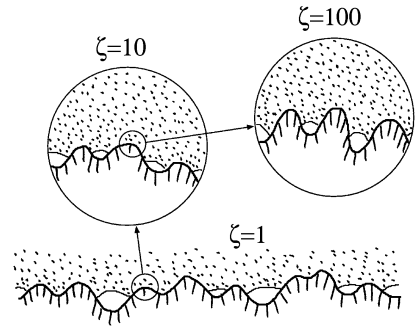} 
\caption{
A rubber block (dotted area) in adhesive contact with a hard
rough substrate (dashed area). The substrate has roughness on many different
length scales and the rubber makes partial contact with the substrate on all length scales.
When a contact area 
is studied at low magnification ($\zeta=1$)
it appears as if complete contact occurs in the macro asperity contact regions,
but when the magnification is increased it is observed that in reality only partial
contact occurs.
}
\label{1x}
\end{figure}

Let us consider the contact between an elastic solid with a flat surface and a hard
randomly rough substrate. 
Fig.~\ref{1x} shows the
contact between the solids at increasing magnification $\zeta$. At low magnification
($\zeta = 1$)
it looks as if complete contact occurs between the solids at many {\it macro asperity}
contact regions,
but when the magnification is increased smaller length scale roughness is detected,
and it is observed that only partial contact occurs at the asperities.
In fact, if there would be no short distance cut-off the true contact area
would vanish. In reality, however,
a short distance cut-off will always exist since the shortest possible length is
an atomic distance. In many cases the local pressure at asperity contact regions
at high magnification will become so high
that the material yields plastically before reaching the atomic dimension.
In these cases
the size of the real contact area will be determined mainly by the yield stress
of the solid. 

The magnification $\zeta$ refers to some (arbitrary) chosen reference length scale.
This could be, e.g., the lateral size $L$ of the nominal contact area in which case 
$\zeta= L/\lambda$, where $\lambda$ is the shortest wavelength
roughness which can be resolved at magnification $\zeta$. In this paper we will instead
use the roll-off wavelength $\lambda_0=2\pi /q_0$ as the reference 
length so that $\zeta = \lambda_0/\lambda$.

Recently, a very general contact mechanics theory has been developed which can
be applied to both stationary and sliding contact for viscoelastic solids
(which includes elastic solids as a special case)\cite{Persson_JCP2001}. 
The theory was originally developed
in order to describe rubber friction on rough substrates. 
For elastic solids the theory can also be 
applied when the adhesional interaction is taken into account\cite{P1}. In contrast
to earlier contact mechanics theories, the theory presented in 
Ref. \cite{Persson_JCP2001,P1} is particularly accurate 
close to complete contact, as would be the case for, e.g., rubber
on smooth surfaces. The basic idea behind the theory is to study the contact 
at different magnification. In particular, the theory describes
the change in the stress distribution $P(\sigma, \zeta)$
as the magnifications $\zeta$ increases. Here 
\begin{equation}
  \label{stressdistributiondef}
  P(\sigma, \zeta)=\langle \delta (\sigma - \sigma({\bf x},\zeta))\rangle
\end{equation}
is the
stress distribution at the interface when the surface roughness with wavelength
smaller than $\lambda = \lambda_0/\zeta$ has been removed. 
In (\ref{stressdistributiondef}), $\langle \ldots \rangle$ stands for ensemble
average, and $\sigma({\bf x},\zeta)$
is the perpendicular stress at the interface when surface roughness with wavelength 
shorter than $\lambda=\lambda_0/\zeta$ has been removed.
It is clear that as the
magnification $\zeta$ increases, the distribution $P(\sigma, \zeta)$ will be broader
and broader and the theory describes this in detail. The (normalized) area of real contact
(projected on the $xy$-plane) at the magnification $\zeta$ can be written as
\begin{equation}
{A(\zeta) \over A_0} = \int_{0^+}^\infty d\sigma \ P(\sigma,\zeta).
\end{equation}
where the lower integration limit $0^+$ indicate that the delta function at the origin
$\sigma = 0$ (arising from the non-contact area) should be excluded from the integral.
The rubber friction theory described in Ref. \cite{Persson_JCP2001} depends on the 
function $A(\zeta )/A_0$ for {\it all
magnifications}. This just reflects the
fact that the friction force results from the viscoelastic deformations of the rubber on all length
scales, and when evaluating the contribution to the friction from the
viscoelastic deformations on the length scale $\lambda$,
it is necessary to know the contact between the rubber
and the substrate at the magnification $\zeta=\lambda_0/\lambda$.
Thus, not just the area of real (atomic) contact is of great interest, but many important applications
require the whole function $A(\zeta)$, and the pressure distribution $P(\sigma, \zeta)$.

In order to accurately reproduce
the contact mechanics between elastic solids, it is in general
necessary to consider
solid blocks which extend a similar distance in the direction normal to
the nominal contact area as the linear size of the contact area. 
This leads to an enormous number of atoms or dynamical variables already
for relatively small systems. In this paper we develop a multiscale approach
to contact mechanics where the number of dynamical variables scales like $\sim N^2$
rather than as $\sim N^3$, where $N\times N$ is the number of atoms in the nominal
contact area. As application we consider the contact mechanics between randomly rough
surfaces both with and without adhesion, and compare the results with
analytical contact mechanics theories.

\vskip 0.5cm
{\bf 2. Multiscale molecular dynamics}

Let us discuss the minimum block-size necessary in a computer simulation
for an accurate description of the
contact mechanics between two semi-infinite elastic solids with nominal flat surfaces. 
Assume that the surface roughness power spectrum has a roll-off
wavevector $q=q_0$ corresponding to the roll-off wavelength $\lambda_0 = 2 \pi /q_0$. 
In this case the minimum block must extend $L_x \approx \lambda_0$
and $L_y \approx \lambda_0$ along the $x$ and $y$-directions.
Furthermore, the block must extend at least a distance $L_z \approx \lambda_0$ 
in the direction 
perpendicular to the nominal contact area.
The latter follows from the fact that a periodic stress distribution
with wavelength $\lambda$ acting on the surface of a semi-infinite elastic solid
gives rise to a deformation field which extends a distance $\sim \lambda$ into the solid.
Thus, the minimum block is a cube with the side $L=\lambda_0$.

\begin{figure}
  \includegraphics[width=0.4\textwidth]{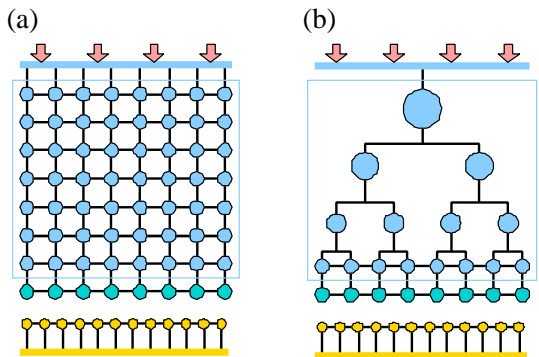} 
  \caption{\label{smartblock}
          Schematic structure of the model. (a) The fully atomistic model.
          (b) The multiscale {\it smartblock} model, 
          where 
          the solid in (a) is coarse grained 
          by replacing groups of atoms with bigger ``atoms''.
          }
\end{figure}

As an example, if $\lambda_0$ corresponds to 1000 atomic spacings,
one must at least consider a block with $1000\times 1000$ atoms
within the $xy$-contact plane, i.e., one would need to study the elastic deformation
in a cubic block with at least $10^9$ atoms.
However, it is possible to drastically reduce the number of 
dynamical variables without loss of accuracy if one notes that an interfacial
roughness with wavelength $\lambda$ will give rise to a deformation field in the block which
extends a distance $\lambda$ into the solid, and which varies spatially over a typical
length scale $\lambda$. Thus when we study the deformation a distance $z$ into the block
we do not need to describe the solid on the atomistic level, but we can coarse-grain 
the solid by replacing groups of atoms with bigger ``atoms'' as indicated schematically
in Fig. \ref{smartblock}. 
If there are $N\times N$ atoms in the nominal contact area
one need $n\approx {\rm ln} N$ ``atomic'' layers in the $z$-direction.
Moreover the number of atoms in each layer decreases in a geometric progression
every time the coarse graining procedure is applied, so that the total number of
particles is of order $N^2$ instead of $N^3$. This results in a huge reduction in the
computation time for large systems.
This multiscale approach may be implemented in various ways, 
and in the Appendix A we outline the procedure
we have used in this paper
which we refer to as the  {\it{smartblock}}. Another implementation
similar to our approach can be found in Ref.~\cite{curtarolo}.

The smartblock model should accurately describe the deformations in the
solids as long as the deformations varies slowly enough with time. However,
the model cannot accurately describe the propagation of short wavelength phonons.
This is, in fact, true with all forms of Hamiltonian multiscale descriptions of
solids, because of the energy conservation together and the unavoidable loss
of information in the coarse grained region.
In principle it should be possible to prevent the back reflection of short
wavelength phonons by describing the coarse grained region as a continuum,
where the numerical calculation can be carried on through a Finite Element
scheme.\cite{NieChenRobbins,E,curtin,cai}
This indeed would require no coarse graining at all in
the region treated with molecular dynamics, and a proper choice of the
matching conditions between the atomistic and the continuum region.
However, with respect to contact mechanics and adhesion the back reflection
of short wavelength phonons is not an important limitation.
With respect to sliding friction it may be a more severe limitation 
in some cases.

Figure \ref{smartblock} illustrates a case where the block is in the form of a cube
with atomically flat surfaces. It is possible to obtain curved surfaces of
nearly arbitrary shape by ``gluing''
the upper surface of the block to a hard curved surface profile. This was described in detail
in Ref. \cite{Persson_JCP2001}. The elastic modulus and the shear modulus of the solid can be fixed at
any value by proper choice of the elongation and bending spring constants for the springs
between the atoms
(see Ref. \cite{Persson_JCP2001} and Appendix A). 
The upper surface of the smartblock can be moved with arbitrary
velocity in any direction, or an external force of arbitrary magnitude can be applied
to the upper surface of the smartblock. We have also studied sliding friction problems where
the upper surface of the smartblock is connected to a spring which is pulled in some prescribed way.
The computer code also allows for various lubricant fluids between the solid 
surfaces of the block and the substrate.
Thus the present model is extremely flexible and can be used to study many interesting
adhesion and friction phenomena, which we will report on elsewhere.

We note that with respect to contact mechanics,
when the slopes of the surfaces are small, i.e. when the surfaces are 
almost horizontal, one of the two surfaces
can be considered flat, while the profile of the other surface has to be 
replaced by the difference of the two original profiles\cite{Jon}. Thus, if the
substrate has the profile $z=h_1({\bf x})$ and the block has the profile
$z=h_2({\bf x})$, then we can replace the actual system with a fictive
one where the block has an atomically smooth surface while the substrate
profile $ h({\bf x}) = h_2({\bf x})-h_1({\bf x})$. Furthermore, if the original
solids have the elastic modulus $E_1$ and $E_2$, and the Poisson ratio $\nu_1$ and
$\nu_2$, then the substrate in the fictive system can be treated as
rigid and the block as elastic with the elastic modulus $E$ and Poisson
ratio $\nu$ chosen so that $(1-\nu^2)/E = (1-\nu_1^2)/E_1+(1-\nu_2^2)/E_2$.

The results presented below have been obtained for a rigid and rough substrate.
The atoms in the bottom layer of the block form a simple square lattice
with lattice constant $a$. The lateral dimensions $L_x=N_xa$ and $L_y=N_ya$. 
For the block, $N_x=400$ and $N_y=400$. Periodic boundary conditions are
applied in the $xy$ plane.  
The lateral size of the block is equal to that of substrate,
but we use different lattice constant $b \approx a/\phi$, where $\phi=(1+\sqrt{5})/2$
is the golden mean, in order to avoid the formation of commensurate
structures at the interface.
The mass of a block atom is 197 a.m.u.
and the lattice constant of the block is $a=2.6~\mbox{\AA}$, reproducing
the atomic mass and the density of gold.
We consider solid blocks with two different Young's moduli: a hard
solid with $E=77 \ {\rm GPa}$, like in gold, and a soft one with
$0.5 \ {\rm GPa}$. The corresponding shear moduli were
$G=27 \ {\rm GPa}$ and $0.18 \ {\rm GPa}$, respectively.

The atoms at the interface between the block and the substrate interact with 
the potential
\begin{equation}
 \label{potential}
 U(r)=4 \epsilon \left [ \left ({r_0\over r}\right )^{12}-\alpha 
 \left ({r_0 \over r}\right )^{6} \right]
\end{equation}
where $r$ is the distance between a pair of atoms.
When $\alpha = 1$, Eq. (\ref{potential}) is the standard
Lennard-Jones potential. The parameter $\epsilon$ is the binding energy
between two atoms at the separation $r=2^{1/6} r_0$.
When we study contact mechanics without adhesion 
we put $\alpha = 0$. In the calculations presented below we have used $r_0= 3.28~\mbox{\AA}$
and $\epsilon = 18.6 \ {\rm meV}$, which (when $\alpha = 1$) gives an 
interfacial binding energy (per unit area)\cite{Is}
$\Delta \gamma \approx 4\epsilon/a^2 \approx 11 \ {\rm meV/\mbox{\AA}^2}$.

\vskip 0.5cm
{\bf 3. Self affine fractal surfaces}

\begin{figure}
  \includegraphics[width=0.55\textwidth]{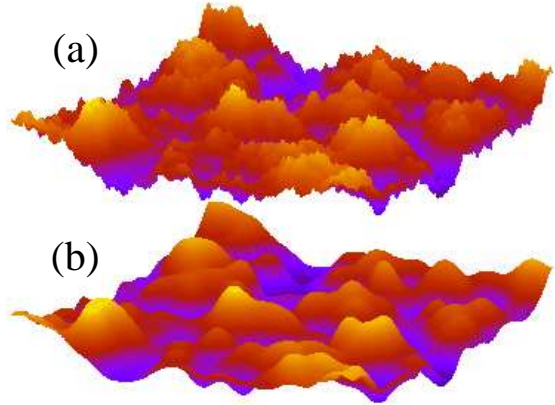} 
  \caption{ \label{fractal_surface}
          (a) Fractal surface with the large wavevector cut-off $q_1= 2\pi /b \approx 216\, q_0$.
          (b) The same surface as in (a) but at lower resolution with $q_1=4q_0$. 
          For a square $1040\,\mbox{\AA}\times 1040\,\mbox{\AA}$ surface area.
          The fractal dimension $D_{\rm f} = 2.2$ and the root-mean-square roughness 
          amplitude is $10~\mbox{\AA}$.}
\end{figure}

In our calculations we have used self affine fractal surfaces generated as outlined in 
Ref. \cite{P3}. 
Thus, the surface height is written as
\begin{equation}
 \label{profile}
 h({\bf x}) =\sum_{\bf q} B({\bf q}) e^{i[{\bf q}\cdot {\bf x}+\phi({\bf q})]}
\end{equation}
where, since $h({\bf x})$ is real, $B(-{\bf q})=B({\bf q})$ and $\phi(-{\bf q})=
-\phi({\bf q})$. If $\phi({\bf q})$ are independent random variables, uniformly distributed
in the interval $[0,2\pi[$, then one can easily show that higher order
correlation functions involving $h({\bf x})$ can be decomposed into a product of pair correlations,
which implies that the height
probability distribution $P_h = \langle \delta (h-h({\bf x}))\rangle $ is Gaussian\cite{P3}.
However, such surfaces can have {\it arbitrary
surface roughness power spectrum}.
To prove this, 
substitute (\ref{profile}) into (\ref{powerspectrum}) and use that
$$\langle e^{i\phi({\bf q'})} e^{i\phi({\bf q''})}\rangle = \delta_{{\bf q'},-{\bf q''}}$$
gives
$$C({\bf q})={1\over (2\pi )^2} \int d^2x \ \sum_{\bf q'} |B({\bf q'})|^2 e^{i({\bf q}-{\bf q'})
\cdot {\bf x}}
$$
$$=
\sum_{{\bf q'}} |B({\bf q'})|^2 \delta ({\bf q}-{\bf q'})$$
Replacing
$$\sum_{{\bf q}} \rightarrow {A_0\over (2\pi)^2}\int d^2q,$$
where $A_0$ is the nominal surface area, gives
$$C({\bf q})={A_0\over (2\pi)^2} |B({\bf q})|^2.$$
Thus, if we choose
\begin{equation}
 \label{Bq}
 B({\bf q})= (2\pi/L) [C({\bf q})]^{1/2},
\end{equation}
where $L=A_0^{1/2}$, then the surface roughness profile (\ref{profile}) has the
surface roughness power density $C({\bf q})$. If we assume that the
statistical properties of the rough surface are isotropic, then $C({\bf q})=C(q)$
is a function of the magnitude $q=|{\bf q}|$, but not of the direction of ${\bf q}$.
The randomly rough substrate surfaces used in our numerical calculations where
generated using (\ref{profile}) and (\ref{Bq}) and assuming that the surface roughness
power spectra have the form shown in Fig.~\ref{Cq1},
with the fractal dimension $D_{\rm f}=2.2$ and 
the roll-off wavevector $q_0=3q_L$, where $q_L= 2\pi /L_x$.
We have chosen $q_0=3q_L$ rather than $q_0=q_L$ since the former value gives 
some self-averaging and less noisy numerical results.
We also used $q_1= 2\pi /b \approx 216q_0$ (topography (a) in 
Fig. \ref{fractal_surface}) and some surfaces with 
several smaller values for $q_1$ (Fig. \ref{fractal_surface} (b) shows the topography when
$q_1=4q_0$), corresponding to lower magnification (see Sec. 4).

\vskip 0.5cm
{\bf 4. Numerical results}

In this section we illustrate our multiscale molecular dynamics (MD) approach
by some applications. We first compare the MD results to
two known contact mechanics results from 
continuum mechanics. 
Next we discuss contact mechanics for randomly rough surfaces
both with and without adhesion.

\vskip 0.2cm
{\bf 4.1. Test cases: Hertz contact and complete contact}

In 1881 Hertz presented an exact solution
for the contact between two perfectly elastic solids with local quadratic profiles.  
The results were derived using the elastic continuum model
and neglecting the adhesion between the solids. In addition, Hertz assumed that
the interfacial friction vanishes so that no shear stress can 
develop at the interface between
the solids. 
When a spherical asperity is squeezed against a flat surface a circular
contact area (radius $r_{\rm H}$) is formed, where the pressure decreases continuously
from the center $r=0$ to the periphery $r=r_{\rm H}$ of the contact according to 
\begin{equation}
 \label{hertzpressure}
 \sigma = \sigma_{\rm H} \left [1-\left ({r\over r_{\rm H}}\right )^2\right ]^{1/2}.
\end{equation}

\begin{figure}
  \includegraphics[width=0.4\textwidth]{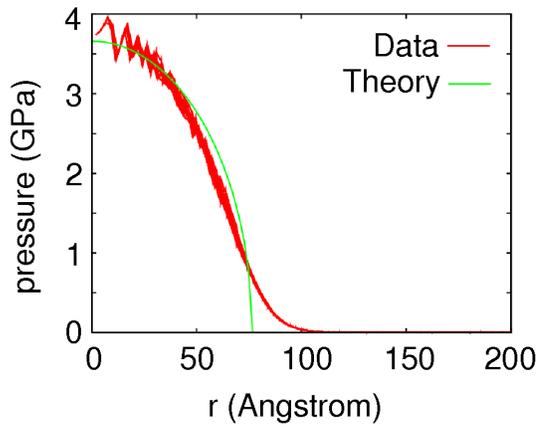} 
  \caption{ \label{pressure_dis_spherical_tip}
    The pressure in the contact region between
    a spherical tip and a flat elastic surface. We show the 
    simulation data and the theoretical Hertz result. The spherical
    tip has the radius of curvature $R=1040~\mbox{\AA}$ and the loading force
    $4.6\times 10^{-7}~\mbox{N}$.}
\end{figure}

Let us compare the prediction of our atomistic model with the Hertz theory.
We use the Lennard-Jones
potential with $\alpha = 0$, i.e. without the attractive term.  
In Fig. 
\ref{pressure_dis_spherical_tip} we compare the Hertz contact pressure (green line)
with our numerical data (red data points).
The numerical data were obtained 
for a rigid spherical tip squeezed against a flat elastic surface.
Note that the pressure obtained from the MD calculation 
has a 
tail beyond the Hertz contact radius $r_{\rm H}$.
Similar ``pressure tails'' were recently observed in 
molecular dynamics simulations by Luan and Robbins\cite{Rob2005}.
The tail reflects the non-zero extent of the atom-atom 
interaction potential. 
The deviation between the molecular dynamics results and the continuum mechanics results should
decrease continuously as the size of the system increases. 

At the atomic level there is no unique way to define when two solids are in contact,
and one may use several different criteria.
One method is based on the force acting between the atoms at the interface
and works best when the adhesional interaction is neglected.
Thus, when two surfaces approach each other,
the repulsive force between the atoms increases. We may define contact when the repulsive force is
above some critical value.
When adhesion is included the interaction between the wall atoms becomes more long-ranged
and it is less obvious how to define contact based on a force criterion,
and we find it more convenient 
to use a criteria based on the nearest neighbor distance between atoms on the two surfaces.
Thus, when the separation between two opposing 
surface atoms is less than some critical value, contact is defined to occur. However,
we have found that neither of these two criteria gives fully satisfactory results.
The reason is that if the critical force or the critical distance used to define 
when contact occurs is determined by fitting the Hertz pressure profile (\ref{hertzpressure}) to the 
numerical data as in Fig. \ref{pressure_dis_spherical_tip}, 
then the resulting values depend on the radius of
curvature of the asperity. 
For example, for the Hertz contact in Fig. \ref{pressure_dis_spherical_tip}
the contact area deduced from the atomistic MD calculation agree with the
Hertz theory if we choose the cut-off pressure $p_{\rm c} \approx 0.7 \ {\rm GPa}$.
However, if the radius of
curvature of the asperity is 10 times smaller ($R=104~\mbox{\AA}$) then, for the same
penetration, the cut-off
would be $p_{\rm c} \approx 2.5 \ {\rm GPa}$, i.e., more than three times larger. 
On the other hand randomly rough surfaces have a wide distribution of
curvatures and it is not clear how to choose the optimum cut-off distance or force.
In this
paper we have therefore used 
another way of determining the contact area which turned out to be more unique. 
We will now describe this method.

\begin{figure}
  \includegraphics[width=0.4\textwidth]{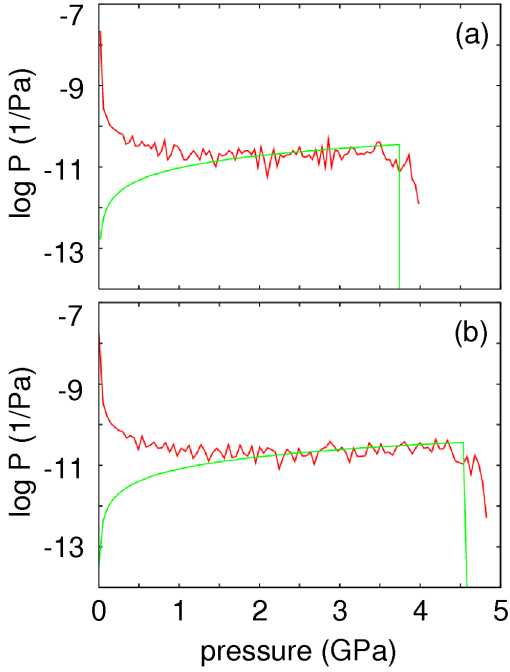} 
  \caption{ \label{pressure.distribution}
    The pressure distribution in the contact region between
    a spherical tip and a flat surface. We show the 
    simulation data (red curves) and the theoretical Hertz 
    result (green curves). Loading force in (a) is $4.6\times 10^{-7} \ {\rm N}$ and in 
    (b) $7.3\times 10^{-7} \ {\rm N}$.}
\end{figure}

Let us consider the pressure distribution $P(\sigma,\zeta )$ at the interface.
For Hertz contact we get the pressure distribution
\begin{equation}
 P(\sigma ) = {1\over A_0} \int_{A_0} d^2x \ \delta (\sigma - \sigma({\bf x}))
\end{equation}
Using $\sigma({\bf x})$ from (\ref{hertzpressure}) for $r<r_{\rm H}$
and $\sigma({\bf x})=0$ for $r> r_{\rm H}$ gives
\begin{equation}
 \label{hertzpressuredist}
 P(\sigma) = \left (1-{A\over A_0}\right )\delta (\sigma)+ 
 {2\sigma \over \sigma_{\rm H}^2} {A\over A_0}
\end{equation}
where $A=\pi r_{\rm H}^2$ is the Hertz contact area.
In Fig. \ref{pressure.distribution} we show 
the pressure distribution in the contact region between
a hard spherical tip and an elastic solid with a flat surface. The 
red curve shows the simulation data, while the green curve is the 
theoretical Hertz result 
obtained by a suitable choice of $A$ in Eq. (\ref{hertzpressuredist}).
Note that while 
the Hertz solution and the atomic MD simulation results
agree very well for large pressure, there is a fundamental difference for small pressure.
Thus, for the Hertz solution, for small 
pressure $\sigma \rightarrow 0$,
$P(\sigma )\sim \sigma $,
while in the atomistic model $P(\sigma )$ increase monotonically as $\sigma \rightarrow 0$.
This difference is due to the long-range interaction between the solid walls
in the atomistic model, which is absent in the Hertz model. When the long range
wall-wall interaction is taken into account the delta function at $\sigma = 0$ in the Hertz
solution (\ref{hertzpressuredist}) will broaden, resulting in a $P(\sigma)$
which (for the small systems considered here)
will decay monotonically with increasing
$\sigma$ as observed for the atomistic model. Note that this effect is of exactly the same 
origin as the ``pressure tail'' for $r > r_{\rm H}$  
in Fig. \ref{pressure_dis_spherical_tip}. 

The fact that $P(\sigma, \zeta)$ vanish linearly with $\sigma$ as $\sigma \rightarrow 0$
is an exact result in continuum mechanics with contact interaction (no long range wall-wall
interaction), and is valid not just for the Hertz contact case, but holds in general
\cite{Persson_PRB_2002}. However, as explained above, 
this
effect will never be observed in the atomistic model if the wall-wall interaction is long-ranged.

Note that the contact area $A$ can be determined directly by fitting the analytical
expression for $P(\sigma )$ for the Hertz contact (Eq. (\ref{hertzpressuredist})) 
to the numerical MD results for large enough pressures
(see Fig. \ref{pressure.distribution}). In the present case, 
for $F_{\rm N} = 4.6\times 10^{-7} \ {\rm N}$ 
(Fig. \ref{pressure.distribution}(a))  
this gives a contact area
$A=\pi r_{\rm H}^2$ which is nearly
identical 
to the one deduced from the fit in Fig. \ref{pressure_dis_spherical_tip}.
A similar procedure can be used to determine the contact area between randomly rough surfaces
using the following analytical expression derived from the contact mechanics theory
of Persson (see Eq. (\ref{pressuredist}) below): 
$$P(\sigma, \zeta) = {1\over 2 (\pi G)^{1/2}} \left (e^{-(\sigma -\sigma_0)^2/4G}
-e^{-(\sigma +\sigma_0)^2/4G}\right ),$$
where $\sigma_0$ is the nominal contact stress, and where the fitting parameter
$G=G(\zeta)$ can be related to the contact area using Eq. (3).
Thus, if $A/A_0 \ll 1$ we have $G=(\sigma_0^2/\pi)(A/A_0)^{-2}$.
We have found (see below) that this expression for $P(\sigma, \zeta)$
can fit the numerical MD data very well (lending support for the
accuracy of the Persson theory), and we have used this method to determine
the contact area as a function of the squeezing force for randomly rough substrates.

\begin{figure}
  \includegraphics[width=0.4\textwidth]{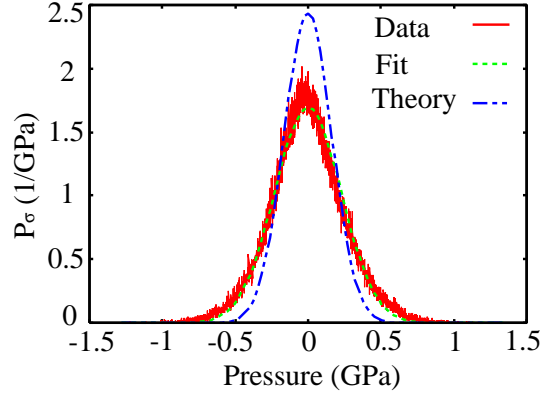} 
  \caption{\label{pressure3}
          The normalized pressure distribution $P(\sigma )$ at the interface
          between an elastic block (elastic modulus $E=0.5 \ {\rm GPa}$) with a flat surface and 
          a rigid randomly rough substrate. Because of adhesion complete contact occurs at the interface.
          The red 
          curve is the simulation result and the green line is the Gaussian fit to 
          the simulation data with the root-mean-square width $\rmssigma=0.229 \ {\rm GPa}$.
          The blue line is the theoretical Gaussian distribution obtained using continuum mechanics
          (see Appendix B). The theoretical 
          rms width $\rmssigma=0.164 \ {\rm GPa}$.
          }
\end{figure}

Let us consider the pressure distribution at the interface between a rigid randomly rough
substrate and a flat elastic surface when the solids are in complete contact. Complete contact
can result either by squeezing the solids together by high enough force, or if the adhesional
interaction between the solids is high enough (or the elastic modulus small enough). However,
when complete contact occurs the pressure distribution is the same. 

For an elastic solid with a flat surface in perfect
contact with a hard randomly rough surface, 
continuum mechanics predict a Gaussian pressure distribution
of the form (see Appendix B):
$$P(\sigma ) = {1\over (2\pi )^{1/2}\rmssigma}e^{-(\sigma-\sigma_0)^2/2\rmssigma^2}$$
where the root-mean-square width $\rmssigma$ is determined by the power spectrum:
$$\rmssigma^2 = \langle \sigma^2\rangle = {\pi \over 2}{E^2\over (1-\nu^2)^2} 
\int_{q_0}^{q_1} dq \ q^3 C(q)$$ 
In Fig. \ref{pressure3} we compare the theoretical pressure distribution (blue curve) with the pressure
distribution obtained from the atomistic model for the case where the
complete contact results from the adhesive interaction between the solids. 
The MD data are well fitted by a Gaussian curve, but the width of the
curve is slightly larger than expected from the continuum mechanics theory
$\rmssigma ({\rm MD})= 0.229 \ {\rm GPa}$ while $\rmssigma ({\rm theory}) = 0.164 \ {\rm GPa}$.
The randomly rough surface used in the MD calculation is self affine fractal the whole way down
to the atomic distance, and one can therefore not expect the continuum mechanics result for
$P(\sigma )$, which assumes ``smooth'' surface roughness, to agree perfectly with the MD result.

\begin{figure}
  \includegraphics[width=0.4\textwidth]{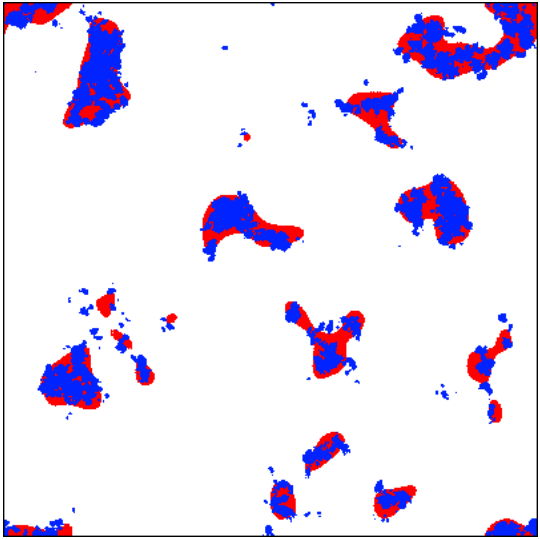} 
  \caption{\label{morph1}
          Contact morphology for two different magnifications.
          The red color denotes 
          contact regions for 
          the low magnification $\zeta=4$, while the blue color corresponds
          to the contact regions for the high magnification $\zeta=216$.
          }
\end{figure}

\vskip 0.2cm
{\bf 4.2. Contact mechanics without adhesion}

Here we study contact mechanics without adhesion as obtained with
$\alpha=0$ in Eq. (4), corresponding to purely repulsive interaction between
the walls.
Fig. \ref{morph1} shows the contact morphologies at different magnifications
$\zeta$ for the same load.
The red and blue color indicate the contact area at low ($\zeta = 4$) and high 
($\zeta = 216$) magnification,
respectively. Note that 
with increasing magnification
the contact area decreases, 
and the 
boundary line of the contact islands becomes rougher. 
In Ref. \cite{Borri-Brunetto_2001} and \cite{PRE_2004} it has been shown that
the statistical properties of the contact regions 
exhibit power-law scaling behavior. 
At low magnification $(\zeta=4)$ it looks as if complete contact
occurs between the solids at asperity contact regions. However, when
the magnification is increased, smaller length scale roughness is detected and
it is observed that only partial contact occurs at the asperities. In fact,
if there were no short distance cut-off in the surface roughness,
the true contact area would eventually
vanish. But in reality a short distance cut-off always exists,
e.g. the interatomic distance.

\begin{figure}
  \includegraphics[width=0.4\textwidth]{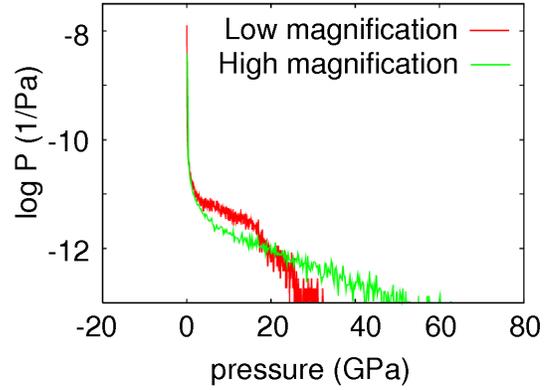} 
  \caption{\label{pressure1}
          The pressure distribution in the contact area for two 
          different magnifications.
          The red line corresponds to the pressure distribution for low
          magnification $\zeta=4$,  
          while the green line is 
          for high magnification $\zeta=216$.
          }
\end{figure}

Fig. \ref{pressure1} shows the pressure distribution in the contact 
area for two different magnifications. When we study contact on shorter 
and shorter length scale, which corresponds to increasing magnification 
$\zeta$, the pressure
distribution becomes broader and broader.

\begin{figure}
  \includegraphics[width=0.4\textwidth]{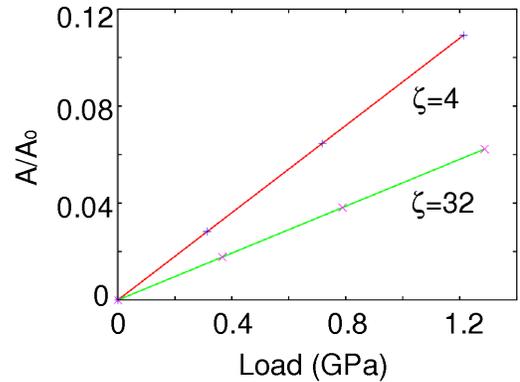} 
  \caption{\label{contactarea_vs_load}
          The relative contact area $A/A_0$, as a function of applied stress
          $F_{\rm N}/A_0$. Results are presented for two different magnifications 
          $\zeta=\lambda_0 /\lambda = 4$ and 32. 
          The fractal dimension is $D_f=2.2$.}
\end{figure}

\begin{figure}
  \includegraphics[width=0.28\textwidth]{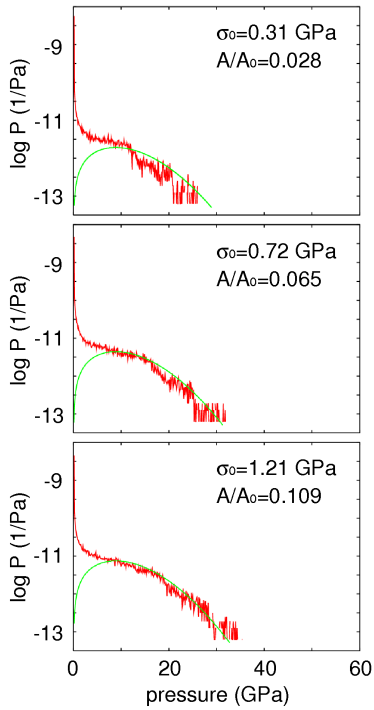} 
  \caption{\label{out4}
          The stress distribution for $\zeta=4$ for three different
          nominal pressure.
          }
\end{figure}

\begin{figure}
  \includegraphics[width=0.28\textwidth]{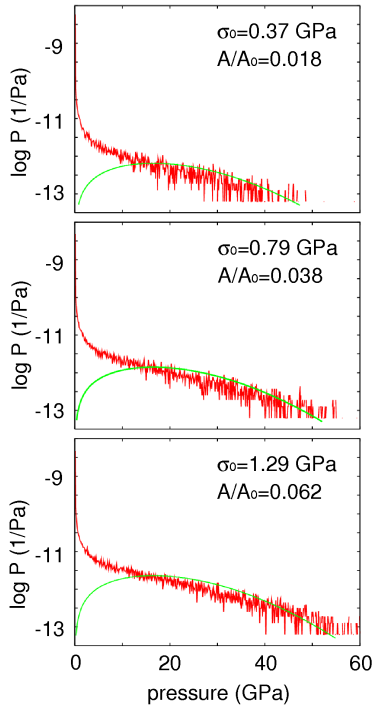} 
  \caption{\label{out32}
          The stress distribution for $\zeta=32$ for three different 
          nominal pressure.
          }
\end{figure}

Fig. \ref{contactarea_vs_load} shows that the contact area varies (approximately) linearly
with the load for the small load at two different magnifications $\zeta = 4$ and $32$.
The contact area was determined as described in Sec. 4.1. by fitting the pressure distribution
to a function of the form (\ref{pressuredist}). The pressure distributions and the fitting functions are shown in Fig.
\ref{out4} and \ref{out32} for $\zeta =4$ and $32$, respectively. The slope of the lines
in Fig. \ref{contactarea_vs_load} is only a factor 1.14 larger than predicted by the contact theory of Persson
(see Sec. 5).

\begin{figure}
  \includegraphics[width=0.4\textwidth]{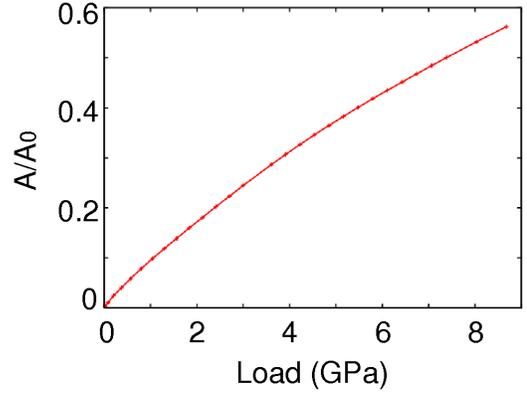} 
  \caption{\label{slope}
          The relative contact area $A/A_0$, as a function of applied stress
          $F_{\rm N}/A_0$. Results are presented
          for the highest magnification $\zeta=216$. Contact is defined when
          the separation between the surfaces is below a critical value. 
          The fractal dimension is $D_f=2.2$.}
\end{figure}

In Fig. \ref{slope} we show the variation of the contact area with the nominal squeezing pressure 
for the highest magnification case $\zeta=216$. In this case we have defined contact to occur when the separation between the surfaces is below some critical value $r_c=4.3615~\mbox{\AA}$. 
In contrast to the definition used above, this
definition does not give a strict linear dependence of the contact area on the load for small load
as found above 
when the contact area is defined using the stress distribution.

\vskip 0.2cm
{\bf 4.3. Contact mechanics with adhesion}

\begin{figure}
  \includegraphics[width=0.4\textwidth]{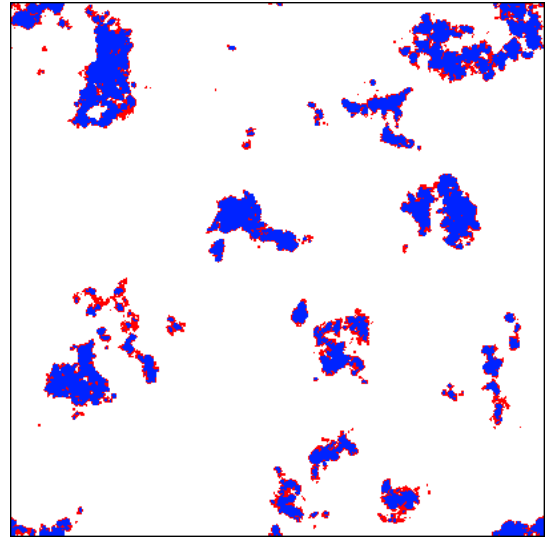} 
  \caption{\label{morph2}
          Contact morphology with adhesion and 
          without adhesion. The blue color region denotes the
          contact without adhesion. The red color region denote the {\it additional} 
          contact area when the adhesional interaction is included.
          }
\end{figure}

In this section we include the adhesive interaction
i.e. we put $\alpha = 1$ in Eq. (4).
Fig. \ref{morph2} presents the contact morphology both with and without the adhesion
at the highest magnification ($\zeta=216$). The regions with blue color denotes the
contact area without adhesion. 
The red color region denotes the {\it additional} 
contact area when adhesion is included.
The contact area with adhesion is, of course, larger than that
without adhesion since the attractive adhesional interaction
will effectively increase the external load\cite{JKR,Ken,Full}.

\begin{figure}
  \includegraphics[width=0.4\textwidth]{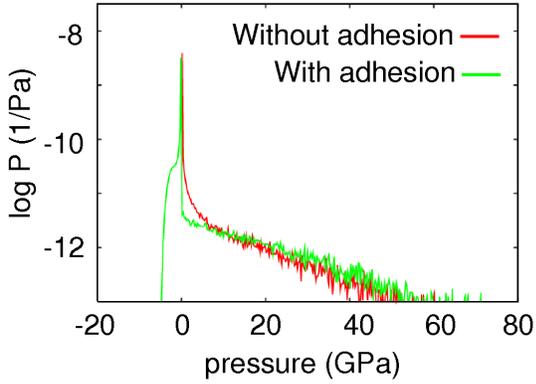} 
  \caption{\label{pressure2}
          The pressure distribution with  
          and without adhesion. The red curve denotes the 
          pressure distribution with adhesion while the green curve 
          is without adhesion.
          }
\end{figure}

Fig. \ref{pressure2} shows the pressure distribution $P(\sigma, \zeta )$ at high
magnification with and without adhesion. When adhesion is 
neglected (corresponding to the $\alpha=0$ in (4)), 
the pressure is positive in the contact area and $P(\sigma,\zeta ) =0$ for $\sigma < 0$.
When the adhesive interaction is included, the stress becomes tensile close to the edges of 
every contact region and $P(\sigma,\zeta )$ is in general finite also for $\sigma < 0$. 

\vskip 0.5cm
{\bf 5. Discussion}

Several analytical theories, based on continuum mechanics, have been developed
to describe the contact between elastic bodies both with and without the
adhesional interaction. Here we will compare the results presented above with
the predictions of some of these theories.

Persson\cite{Persson_JCP2001,P1} has developed a contact mechanics 
theory where the surfaces are studied at different
magnification $\zeta = \lambda_0/\lambda$, where $\lambda_0$ is the roll-off wavelength 
and $\lambda$
the shortest wavelength roughness which can be observed at the magnification $\zeta$.
In this theory\cite{Persson_JCP2001} the stress distribution 
$P(\sigma, \zeta)$ at the interface between the
block and the substrate has been shown to obey (approximately) a diffusion-like
equation where time is replaced by magnification and spatial coordinate by the stress $\sigma$.
When the magnification is so small that no atomic structure can be detected, the surface
roughness will be smooth (no abrupt or step-like changes in the height profile) and one can
then show\cite{Persson_PRB_2002} that in the absence of 
adhesion $P(0,\zeta)=0$. Using this boundary condition  
the solution to the diffusion-like equation gives the pressure distribution 
at the
interface ($\sigma > 0$):
\begin{equation}
 \label{pressuredist}
 P(\sigma, \zeta) = {1\over 2 (\pi G)^{1/2}} \left (e^{-(\sigma -\sigma_0)^2/4G}
 -e^{-(\sigma +\sigma_0)^2/4G}\right )
\end{equation}
where 
\begin{equation}
 \label{G}
 G= {\pi \over 4} \left ( {E\over 1 - \nu^2}\right )^2 \int_{q_L}^{\zeta q_0} 
 dq \ q^3 C(q) \,.
\end{equation}
The relative contact area
\begin{equation}
 \label{relatcontact1}
 {A\over A_0} = \int_0^\infty  d\sigma \ P(\sigma,\zeta) \,.
\end{equation}
Substituting (\ref{pressuredist}) into (\ref{relatcontact1}) gives after some simplifications
\begin{equation}
 \label{relatcontact2}
 {A\over A_0} =
 {1\over (\pi G)^{1/2}} \int_0^{\sigma_0} d\sigma \ e^{-\sigma^2/4G} .
\end{equation}
Thus, for small nominal squeezing pressure $\sigma_0 \ll G^{1/2}$ we get
\begin{equation}
 \label{relatcontact3}
 {A\over A_0} = {\sigma_0 \over (\pi G)^{1/2}} \,.
\end{equation}
Since the squeezing force $F_{\rm N} = \sigma_0 A_0$ we can also write
\begin{equation}
 \label{contactarea}
 A=\kappa\frac{F_{\rm N}}{E^{*}}
 \left( \int d^{2}q \ q^{2}C(q)\right) ^{-1/2}
\end{equation}
where $E^{*}=E/(1-\nu^2)$ and $\kappa=(8/\pi)^{1/2}$. Thus, for small squeezing force
$F_{\rm N}$ the theory predicts a linear dependence of the area of real contact on the load.

For very high squeezing force $\sigma_0 \gg G^{1/2}$ complete contact will occur at the
interface. In this case the second term on the rhs in (\ref{pressuredist}) can be neglected, 
so the pressure distribution is a Gaussian centered at $\sigma_0$ and with
the root-mean-square width $\rmssigma=(2G)^{1/2}$. This result is exact (see Appendix B).
Thus, the theory of Persson is expected to
give a good description of the contact mechanics for all squeezing forces. 
All other analytical contact
mechanics theories are only valid when the squeezing force is so small that the area of
real contact is (nearly) proportional to $F_{\rm N}$. 
But in many important applications, e.g.,
in the context of rubber friction and rubber adhesion, 
the area of real contact for smooth surfaces is often 
close to the nominal contact area.
 
The standard theory of Greenwood and Williamson
\cite{GreenW} describe the contact between rough surfaces 
(in the absence of adhesion), where the asperities are approximated by 
spherical cups with equal radius of curvature
but with Gaussian distributed heights. 
In this theory the area of real contact dependent (slightly) non-linearly on the load
for small load, and can therefore not be directly compared with the Persson result (\ref{contactarea}). 
Bush et al \cite{Bush} developed a more general and accurate contact theory.
They assumed that the rough surface consists of a mean plane with hills and
valleys randomly distributed on it. The summits of these hills are approximated
by paraboloids, whose distributions and principal curvatures are obtained from
the random precess theory. 
As a result of more random nature of the surface, Bush et al found 
that at small load the area of contact depends linearly on the load according
to (\ref{contactarea}) but with $\kappa = (2 \pi )^{1/2}$.
Thus the contact
area of Persson's theory is a factor of $2/\pi$ smaller than that predicted
by Bush. Both the theory of Greenwood and Williamson and 
the theory of Bush et al assume that the asperity contact regions are 
independent. 
However, as discussed in \cite{Persson_PRB_2002}, for real surfaces
(which always have surface roughness on many different length scales) this 
will never be the case even at a very low nominal contact pressure, which 
may be the origin of difference of $2/\pi$ between Persson's theory and Bush's
theory.

Hyun et al performed a finite-element analysis of contact between elastic
self-affine fractal surfaces
\cite{PRE_2004}. 
The simulations were done for rough elastic 
surface contacting a perfectly rigid flat surface.
They found that the contact area varies linearly with the load for small load.
The factor $\kappa$ was 
found to be between the results of the Bush and Persson
theories for all fractal dimensions $D_{\rm f}$.
For $D_{\rm f}=2.2$ (corresponding to $H=0.8$) they found
that $\kappa$ was only $\sim 13\%$ larger than predicted by the Persson theory.

\begin{figure}
  \includegraphics[width=0.28\textwidth]{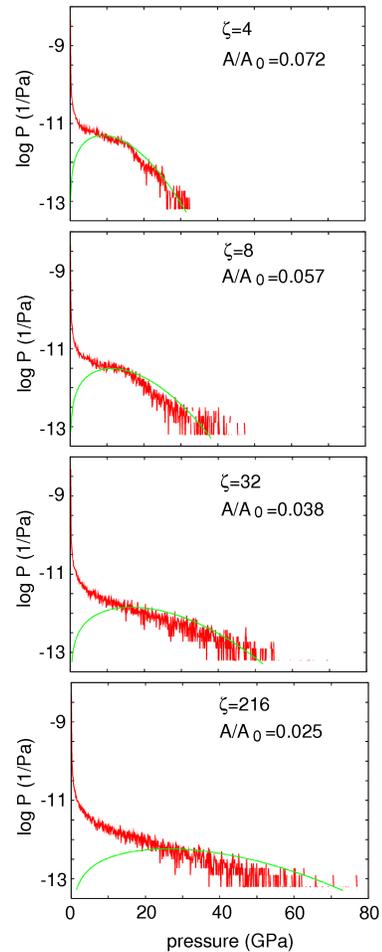} 
  \caption{\label{magi}
          The pressure distribution at four different magnifications
          $\zeta=q_1/q_0 =4$, 8, 32 and 216 for the squeezing pressure
          $\sigma_0 = 800 \ {\rm MPa}$.   
          The red curves is the  
          pressure distribution obtained from the computer simulation,
          while the green curves is from the analytical theory assuming that
          $G^{-1/2}$, and hence
          the relative contact area, is a factor of 1.14 larger than predicted by the
          analytical theory, Eq. (10).} 
\end{figure}

The red curves in Fig. \ref{magi} shows the pressure distribution 
from the simulations for several different values of the magnification
$\zeta = q_1/q_0 = 4$, 8, 32 and 216, neglecting the adhesion. 
In the simulations 
the nominal squeezing pressure $\sigma_0=800 \ {\rm MPa}$.
The best fit (green curves in Fig. \ref{magi})
of the pressure distribution (\ref{pressuredist}) to the numerical results is obtained if 
$G^{-1/2}$ is taken to be a factor 1.14 larger than predicted by the Persson theory
[Eq. (10)], corresponding to a contact area which is
$14\%$ larger than predicted by the analytical theory, in good agreement with the results
obtained by Hyun et al. 

Our simulations show that the contact
area varies linearly with the load for small load,
see Fig. \ref{contactarea_vs_load}. Figs. \ref{contactarea_vs_load} and \ref{magi} show that the slope 
$\alpha(\zeta)$ of the line $A=\alpha(\zeta)F$ decreases with increasing 
magnification $\zeta$, as predicted by the analytical theory 
\cite{Persson_PRB_2002,Borri-Brunetto_2001}. Thus, while $A/A_0=0.072$ for
$\zeta = 4$ we get $A/A_0=0.038$ for $\zeta = 32$, which both are $14\%$ larger than predicted
by Eq. (\ref{relatcontact2}).

\vskip 0.5cm
{\bf 6. Summary and conclusion}

In this paper we have developed a Molecular Dynamics 
multiscale model, which we have used to study the contact between 
surfaces
which are rough on 
many different length scales. We have studied the contact morphologies 
both at 
high and low magnification, with and without adhesion. 
We have shown that in atomistic models it is a non-trivial problem how to define the 
area of real contact between two solids.
Our study shows that the area of real contact is best defined by studying the
interfacial pressure distribution, and fitting it to an analytical expression.
The numerical results are consistent with the theoretical results 
that the contact area varies
linearly with the load for small load, where the proportionality constant 
depends on the magnification $L/\lambda$. For a randomly rough surfaces
with the fractal dimension $D_{\rm f} = 2.2$ (which is typical for many real
surfaces, e.g., produced by fracture or by blasting with small particles) we
have found that for small load (where the contact area is proportional
to the load) the numerical study gives an area of atomic contact
which is only $\sim 14\%$ larger than predicted by the
analytical theory of Persson. Since the Persson's theory is exact in the limit of
complete contact, it is likely that the
Persson theory is even better for
higher squeezing loads.

\vskip 0.5cm
{\bf Acknowledgments}
This work was partly sponsored by MIUR FIRB RBAU017S8 R004,
FIRB RBAU01LX5H, MIUR COFIN 2003 and PRIN-COFIN2004

\vskip 0.5cm
{\bf Appendix A: The ``smartblock'' model for multiscale molecular dynamics}

\begin{figure}
\includegraphics[width=0.4\textwidth]{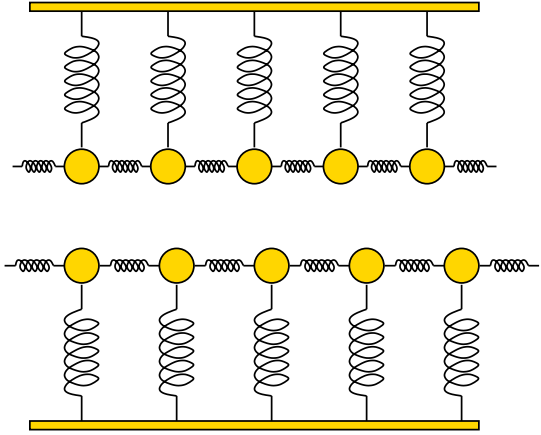} 
  \caption{\label{App1}
The model of Persson and Ballone with long range elasticity.
Side view.}
\end{figure}

Here we present a detailed description of the multiscale model
implemented in our Molecular Dynamics (MD) simulations.
Persson and Ballone\cite{ballone2000} introduced
a simple and effective model to study the boundary
lubrication between elastic walls. For each wall only the outermost
layer of atoms was considered. These atoms were able to interact with
the lubricant or with the atoms of the other wall with Lennard-Jones
potentials. The walls' atoms were connected to a rigid surface through
special springs which exert an elastic reaction not only to elongation,
but also to lateral bending.
The walls' atoms are coupled with their in-plane neighbors with
similar springs. It is also possible to use curved elastic walls by
connecting the vertical springs to a curved rigid surface rather than a flat surface as
in Fig. \ref{App1}. 

The model of Persson and Ballone catches two essential features:
Firstly the walls are not rigid, they can deform (differently from previous models)
and the description takes into account the elastic energy stored during
compression or stretching, which is an essential ingredient for the
study of the squeeze-out. Secondly 
both the Young modulus and the shear modulus can be
independently tuned via the choice of the elastic constants of the
springs.

The model of Persson and Ballone works well when the solid is exposed to uniform shear
or uniform elongation or compression.
However, when there are spatial variations in the stress at
the surface of the solid,
for instance when the displacement at the interface comprises
short wavelength Fourier components, then the model 
does not allow a
proper description of the elastic deformation field. 
In particular, when a periodic stress acts on the surface of an elastic solid,
the displacement field decays exponentially into the solid, and this aspect
is 
absent in the Persson-Ballone model.

The solution to overcome this limitation is straightforward: we
explicitly introduce many layers of atoms, placed on the points of a
simple cubic lattice, and coupled with springs to their nearest
neighbors.

The ``springs'', as in the previous model, are special, since they can
resist to lateral bending. The force due to a vertical spring connecting
two consecutive atoms 1 and 2 along the $z$ axis is given by the
formulas below, where $a$ is the lattice spacing, that is the equilibrium
length of the spring:
$$F_x = - k_b \Delta x = - k_b (x_2-x_1),\eqno(A1)$$
$$F_y = - k_b \Delta y = - k_b (y_2-y_1),\eqno(A2)$$
$$F_z = - k (\Delta z-a) = - k [z_2 - (z_1 +a)].\eqno(A3)$$
Analogous formulas hold for the springs parallel to the $y$ and to the
$z$ axes.
The two elastic constants of the spring, namely $k$ and $k_b$, are
related to the Young modulus $E$ and the shear modulus $G$ respectively:
$k = E a$ and 
$k_b = G a$.

In some circumstances it is useful to simulate
quite large and thick samples. Moreover high resolution up to the atomic
level is needed in part of the sample, typically at the interface.
The solution to avoid excessive computational time is a multiscale
approach: high resolution is achieved where it is needed, but a coarse
grained description is employed when it is feasible.
The coarse graining can happen more times, and to various degrees of
resolution, so that a multilevel description of the system comprising
many hierarchies is implemented.

The grid structure of the smartblock allows a simple procedure to
achieve a multiscale description: groups of atoms can be replaced by
single, bigger atoms, and the elastic constants of the springs are
redefined to guarantee the same elastic response. In many calculations
performed by our group we used to replace a cube of $2\times 2 \times 2$
particles with a single particle, repeating this merging procedure
every two layers.
More generally any change of resolution involves merging together a box
made of $m_x \times m_y \times m_z$ particles. The three numbers $m_x$,
$m_y$ and $m_z$ are called \emph{merging factors} along the three axes.

The equilibrium position of the new particle is in the center of mass
of the group of particles merged together. Its mass is $m_x m_y m_z$
times the mass of the original particles, so that the density does not
change. In fact the masses are only important to study the kinetic, but
they do not influence the static equilibrium configuration.

\begin{figure}
\includegraphics[height=0.4\textheight]{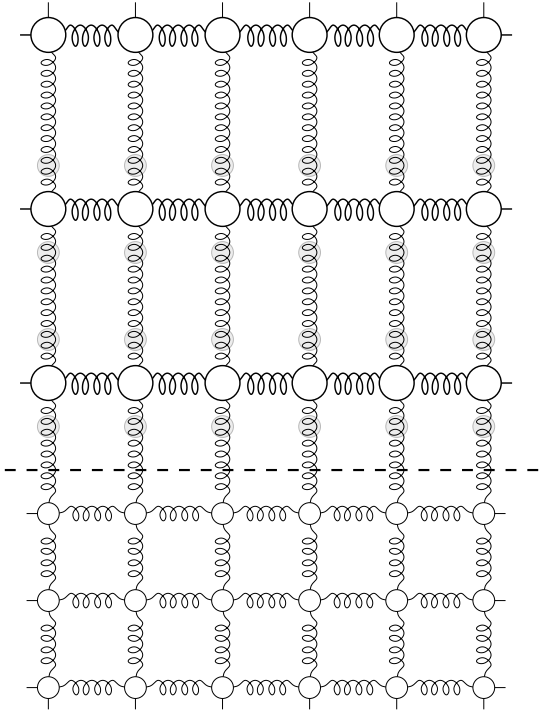} 
\caption{  \label{merging}
The grid of particles is coarse grained by replacing two atoms with a
single one. Merging factors $m_x=m_y=1$, $m_z=2$. Masses, equilibrium
positions and spring constants are changed accordingly.
}
\end{figure}

The three merging factors can be chosen independently. The easiest way
to calculate the new springs' elastic constants is by considering the
merging only along one of the axes. Fig.~\ref{merging} sketches the
case $m_z=2$, $m_x=m_y=1$ (no change of lattice constant along $x$ and
$y$). Along the direction of merging the new spring constants for
elongation and bending are $k'=k/m_z$ and $k_b'=k_b/m_z$ respectively.
The longer springs get proportionally smaller elastic constants, as it
happens when springs are connected in series.
In the two directions orthogonal each spring replaces $m_z$ old springs
in parallel configuration, so the elastic constants increase
proportionally: $k'=m_x k$, $k_b'=m_x k_b$.
Below there is the general formula giving the new elastic constants of
the springs along the $z$ axis, with arbitrary merging factors:
$$k' = \frac{m_x m_y}{m_z} k \,; \ \ \ \ \ \ \    k_b' 
= \frac{m_x m_y}{m_z} k_b \eqno(A4)$$
Analogous formulas hold for the springs parallel to the $x$ and $y$ axes.

To get the whole picture we have to characterize the springs at the
interface between the two lattices, e.g., the ones crossing the dashed
line in Fig.~\ref{merging}. When the merging is in the direction $z$
orthogonal to the interface both elastic constants
$k$ and $k_b$ get multiplied by the factor $2/(1+m_z)$.
Actually their length is $\frac{1}{2}(m_z+1)a_z$, $a_z$ being the old
lattice constant along $z$. Each of these interface springs can be
thought as half a spring of the old grid connected with half a spring
of the new grid.

\begin{figure}
\includegraphics[width=0.4\textwidth]{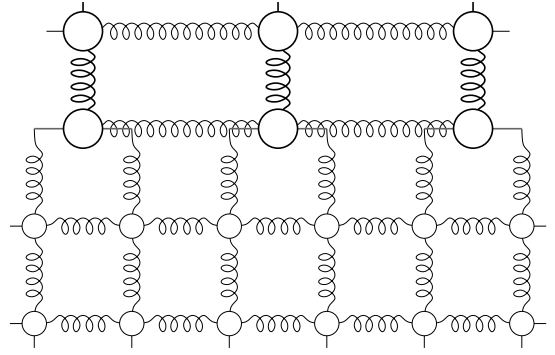} 
\caption{  \label{merge_along_x}
Change of lattice spacing along a direction parallel to the interface
between the two grids.
}
\end{figure}

When the merging is along a direction orthogonal to the interface
between the two grids, as sketched in Fig.~\ref{merge_along_x}, then
the spring constants do not change, but the forces between the particles
are calculated taking into account the in-plane shift between the atoms
of the two grids. Each interface particle of the upper lattice interacts
with $m_x\times m_y$ particles of the lower lattice. The equations
(A1) and (A2) are modified: $F_x=-k_b (\Delta x +
\mbox{x-shift})$,  $F_y=-k_b (\Delta x +\mbox{y-shift})$. The two
in-plane shifts depend on the pair of particles considered.

\vskip 0.5cm
{\bf Appendix B: Pressure distribution at complete contact between 
randomly rough surfaces}

Here we calculate the pressure distribution at the interface between
two solids in complete contact. We assume that one solid is rigid and
randomly rough and the other solid elastic with a flat surface.
The pressure distribution
$$P(\sigma)= \langle \delta (\sigma-\sigma({\bf x}))\rangle = {1\over 2 \pi}
\int_{-\infty}^\infty d\alpha \ \langle e^{i\alpha (\sigma-\sigma({\bf x}))}\rangle$$
$$= {1\over 2 \pi} \int_{-\infty}^\infty d\alpha \ 
e^{i\alpha \sigma} F(\alpha)\eqno(B1)$$
where
$$F(\alpha ) = 
\langle e^{-i\alpha \sigma({\bf x}))}\rangle$$
where $\sigma ({\bf x})$ is the fluctuating pressure at the interface.
Next, writing
$$\sigma({\bf x}) = \int d^2q \ \sigma ({\bf q})e^{i{\bf q}\cdot {\bf x}}$$
$$= \int d^2q \ {Eq\over 2(1-\nu^2)} h({\bf q}) e^{i{\bf q}\cdot {\bf x}}$$
where we have used the relation between $\sigma ({\bf q})$ and the Fourier transform 
$h({\bf q})$ of
the height profile $h({\bf x})$ derived in Ref. \cite{Persson_JCP2001}, we get
$$F=
\left\langle {\rm exp} \left ( -i\alpha 
\int d^2q \ {Eq\over 2(1-\nu^2)} h({\bf q}) e^{i{\bf q}\cdot {\bf x}}\right )
\right\rangle$$
Next, using that $h({\bf q})$ are independent random variables we get
$$F=
e^{-\alpha^2 \xi^2/2}\eqno(B2)$$
where
$$\xi^2 = 
\int d^2q d^2q' \ \left ({E\over 2(1-\nu^2)} 
\right )^2 qq'
\langle h({\bf q}) h({\bf q'})\rangle e^{i({\bf q}+{\bf q'})\cdot {\bf x}}
$$
However (see Ref. \cite{Persson_JCP2001})
$$\langle h({\bf q}) h({\bf q'})\rangle = C(q)\delta ({\bf q}+{\bf q'})$$ 
so that
$$\xi^2 = 
\int d^2q \ \left ( {Eq\over 2(1-\nu^2)}\right )^2 
C(q)
\eqno(B3)$$
Substituting (B2) in (B1) and performing the $\alpha$-integral
and using (B3) gives
$$P(\sigma ) = {1\over (2\pi )^{1/2}\rmssigma}e^{-\sigma^2/2\rmssigma^2}$$
where the root-mean-square width $\rmssigma$ is determined by the power spectrum:
$$\rmssigma^2 = \langle \sigma^2\rangle = {\pi \over 2}{E^2\over (1-\nu^2)^2} 
\int_{q_0}^{q_1} dq \ q^3 C(q)$$

\end{document}